\documentclass[sigconf,nonacm]{acmart}
\usepackage{tabularray}
\begin{document}
\title{StoryNavi: On-Demand Narrative-Driven Reconstruction of Video Play With Generative AI}
\author{Alston Lantian Xu}
\affiliation{%
  \institution{City University of Hong Kong}
  \country{~}
  \vspace{-1.2em}
}
\email{lantian.xu.alston@my.cityu.edu.hk}

\author{Tianwei Ma}
\affiliation{%
  \institution{City University of Hong Kong}
  \country{~}
  \vspace{-1.2em}
}
\email{tianweima_96@gwmail.gwu.edu}

\author{Tianmeng Liu}
\affiliation{%
  \institution{University of Washington}
  \country{~}
  \vspace{-1.2em}
}
\email{tliu45@uw.edu}

\author{Can Liu}
\affiliation{%
  \institution{City University of Hong Kong}
  \country{~}
  \vspace{-1.2em}
}
\email{canliu@cityu.edu.hk}
\authornotemark[1]

\author{Alvaro Cassinelli}
\affiliation{%
  \institution{City University of Hong Kong}
  \country{~}
  \vspace{-1.2em}
}
\email{acassine@cityu.edu.hk}
\authornote{Co-corresponding authors}

\begin{abstract}
Manually navigating lengthy videos to seek information or answer questions can be a tedious and time-consuming task for users. We introduce StoryNavi, a novel system powered by VLLMs for generating customised video play experiences by retrieving materials from original videos. It directly answers users' query by constructing non-linear sequence with identified relevant clips to form a cohesive narrative. StoryNavi offers two modes of playback of the constructed video plays: 1) video-centric, which plays original audio and skips irrelevant segments, and 2) narrative-centric, narration guides the experience, and the original audio is muted. Our technical evaluation showed adequate retrieval performance compared to human retrieval. Our user evaluation shows that maintaining narrative coherence significantly enhances user engagement when viewing disjointed video segments. However, factors like video genre, content, and the query itself may lead to varying user preferences for the playback mode.  
\end{abstract}


\begin{teaserfigure}
  \includegraphics[width=\textwidth]{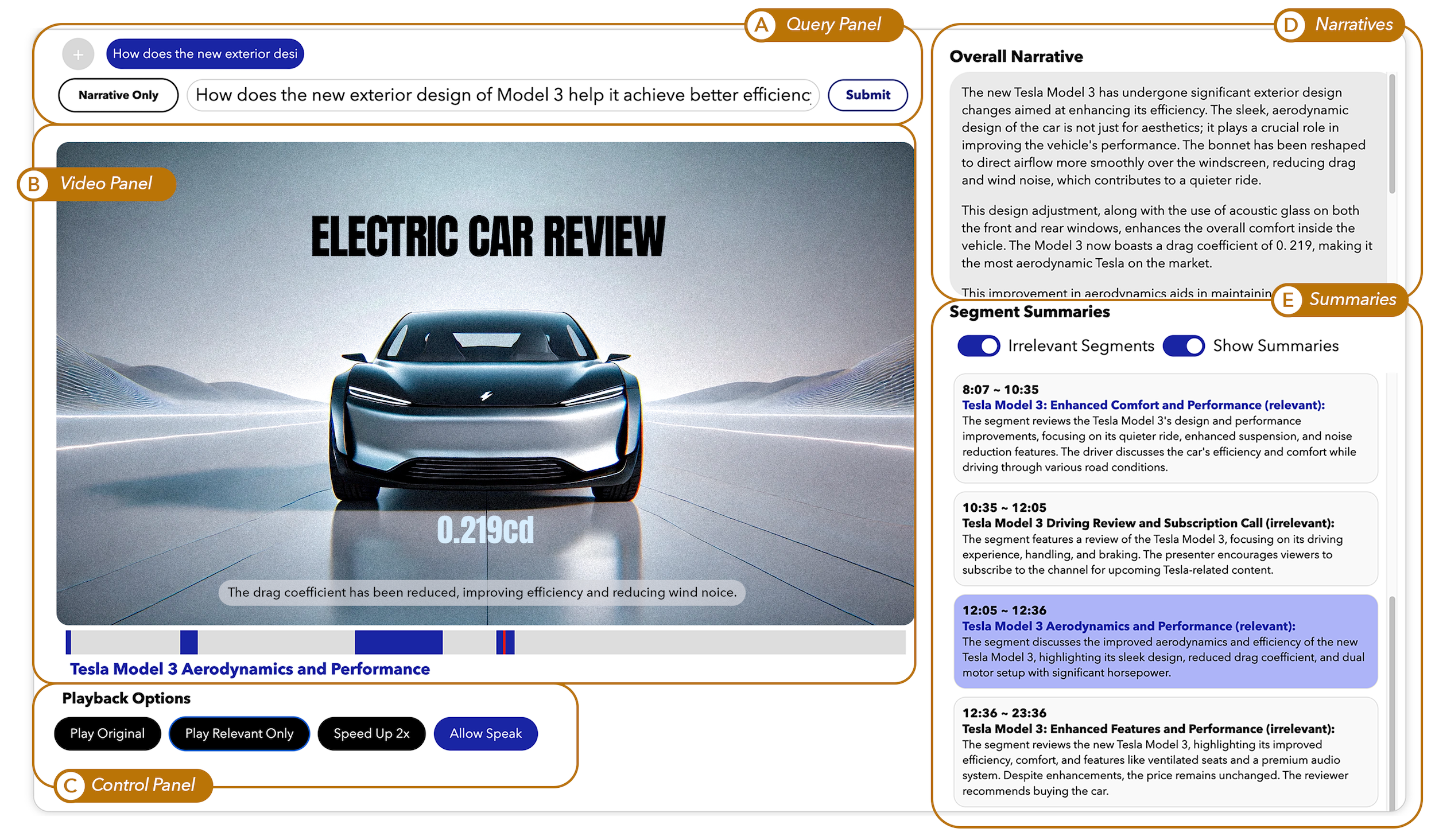}
  \caption{StoryNavi is a query-based, non-linear video navigation system designed to enhance video content retrieval. Retrieved segments would be rearranged according to a cohesive narrative. A) Query Panel: Users input queries to retrieve relevant segments and select the playback mode before submission. Multiple tabs support simultaneous queries. B) Video Panel: The video plays here, with blue blocks below marking relevant segments. A synthesised voiceover appears over the video in Narrative-centric mode. Artificial transition "title cards" appear over the video in Video-centric mode. C) Control Panel: Playback controls allow users to play only relevant segments and adjust playback speed. D) Overall Narrative: An LLM-generated narrative summarizing the retrieved segments. E) Segment Summaries: Titles and summaries for each segment are displayed, with the playing segment highlighted. Note: the video content has been replaced with an AI-generated image due to copyright concerns.}
  \label{fig:teaser}
\end{teaserfigure}

\maketitle

\section{Introduction}
Navigating video content can be a cumbersome and time-consuming experience for users seeking specific information. Traditional video interfaces often rely on linear timelines and direct manipulation (e.g., scrubbing), which can be inefficient for users searching for distinct segments or specific content. As video platforms grow in complexity, the demand for more efficient methods of video retrieval has become crucial. Recent advancements in voice-based video navigation and content-based referencing have addressed some of these inefficiencies, allowing users to interact more naturally with video content. However, these methods typically focus on supporting users' manual temporal manipulation (e.g., jumping to specific timestamps). 

Large language models (LLMs) have been increasingly integrated into everyday applications, simplifying tasks through natural communication. Recently they have also been used for automatically identifying video clips and even editing them. For instance, text-based video editing tools now allow users to select footage based on specific requests\cite{wang_lave_2024,leake_chunkyedit_2024}. However, viewers still face challenges when identifying target clips. Previous works like DemoCut\cite{chi_democut_2013} and OnPause\cite{tuncer2020pause} have explored ways to reduce viewer effort in instructional videos, but they often fail to maintain narrative flow when segments are presented out of sequence. Systems like RubySlippers\cite{chang2021rubyslippers} improve user interaction with content-based referencing but does not support any narrative coherence.

In this paper, we introduce StoryNavi, a novel system that enables automatic identification of video clips based on a user query, and reconstructs in real-time a new video play experience, using clips from the original video and LLM-generated accompanying text. 
Our system leverages vision large language models (VLLMs) to provide a seamless, narrative-driven experience, ensuring that users not only locate the desired content but also understand its context within the broader storyline. Unlike traditional systems that focus on isolated segment retrieval, our approach prioritises the user’s comprehension of the overarching narrative, offering an on-demand customized video-watching experience. This is particularly important in instructional or complex content, where the user may need to connect different video segments into a cohesive understanding of the material.

Our observational study reveals that smooth transitions—both visual and semantic—are key to maintaining narrative coherence. StoryNavi addresses these needs by integrating relevant segments into a unified narrative, offering a more cohesive viewing experience than traditional systems.

We introduce two distinct playback modes for the reconstructed video play: a video-centric mode, which introduces artificial transitions or ``title cards'' between segments, and a narrative-centric mode, where synthesised voiceover (generated as part of the narrative) replaces the original audio. We performed a technical evaluation of our AI pipeline and showed its adequate performance compared with human work. We also conducted a user study comparing two playback modes across a few video types. The user study focused on how each style of narrative presentation affects the user’s understanding of discrete video segments. Results showed that participants preferred the narrative-centric mode for complex scenarios but favoured original audio for straightforward tasks. These findings offer key insights into segment retrieval and presentation design, highlighting implications for the future of narrative-driven video navigation. This paper makes the following main contributions:
 \begin{itemize} 
    \item A novel AI pipeline for extracting and reconstructing video clips maintaining narrative coherence, implemented in a system that supports on-demand non-linear video generation and play in real-time;
    \item Technical evaluation of the pipeline;
    \item Empirical findings about the user experience of narrative-based AI-reconstructed video play and implications for building future systems alike. 
\end{itemize}

\section{Related Work}
Our research builds on advancements in video content understanding and narrative-focused pipelines, enhancing user experiences through structured video segmentation and interactive narrative systems. 

\subsection{Video Annotation and Segmentation}
Efficient extraction of relevant content from videos often relies on annotation, segmentation, and recognition techniques. Manual annotation, while precise, is impractical for large-scale datasets, prompting the adoption of machine learning methods for video segmentation. Techniques such as convolutional neural networks (CNNs) and recurrent neural networks (RNNs) have improved object detection and scene recognition through temporal and contextual analysis\cite{zhou2022survey,8362936,Sibechi_2019_ICCV,8319974,8296302,pinheiro14}.
Recent works, like Video-LLaMA\cite{damonlpsg2023videollama}, have extended these capabilities by integrating causal actions and audio transcription to provide multimodal video \linebreak comprehension\cite{chadha2020iperceive, 6009223, 915358, petkovic2001content}.
Similarly, the "Multi-modal Video Summarisation" model\cite{mmvs2024} simplifies video annotation and retrieval by generating query-dependent summaries based on combined visual and textual data. However, these methods do not directly address user-initiated video navigation, particularly the challenge of maintaining narrative coherence across retrieved segments.

Although our work does not directly contribute to video annotation or segmentation methodologies, we draw upon these principles for segment retrieval. By leveraging user queries, annotations, and transcripts, our system retrieves relevant video segments and constructs a cohesive narrative using LLMs. Unlike multimodal approaches, we focus on high-level concepts rather than low-level details, ensuring that the narrative guides the playback of non-linear video segments, ultimately improving the overall viewing experience.

\subsection{Query-based Interfaces}
Instruction-based video interfaces have significantly improved the ease with which users can access specific information within videos. These interfaces leverage techniques ranging from automated video segmentation to natural language processing (NLP) for seamless user interaction. For example, DemoCut\cite{chi_democut_2013} automatically generates instructional videos by aligning video content with step-by-step text instructions. OnPause\cite{tuncer2020pause} further enhances user interaction by allowing users to pause videos and align their tasks with the video timeline.

More recent systems incorporate LLMs to respond to complex user queries, offering precise and contextually relevant segment retrieval. For instance, tools like ChunkyEdit\cite{leake_chunkyedit_2024} and LAVE\cite{wang_lave_2024} assist editors by grouping raw footage by topic or suggesting edits using LLMs. B-script\cite{huber2019b} recommends b-roll footage for video editors based on transcript with the help of LLM. However, while these systems provide efficient content retrieval for video editing, they primarily target content creators rather than focusing on improving the video-watching experience from the user’s perspective.

\subsection{Interactive Video Narratives}
Maintaining narrative coherence is crucial for user comprehension\cite{yarkoni2008neural}. Early works like DRAGON\cite{DRAGON2008} explored object-path-based frame manipulation for narrative continuity, emphasising the balance between user autonomy and authorial control\cite{steiner2004narrative}. More recent approaches, such as Fried et al.\cite{fried2019text} explored replacing original transcripts and voiceovers to reinforce narrative coherence for video editors, and Rescribe\cite{pavel2020rescribe} focused on refining narrative delivery, particularly for specialised audiences like visually impaired viewers. 

Recent research has also explored how to maintain a coherent narrative across video segments that are far apart in time. Although advancements in accurate segment retrieval have improved non-linear video viewing experiences, the presentation of these segments still requires further attention. The "Branching" approach organises video frames into a tree structure based on narrative or visual elements, facilitating interactive exploration of narrative pathways and personalised user experience \cite{soltani2023design,sheoran2023narrative}. SIVA Suite\cite{siva2015} exemplifies this by enabling free navigation among branches and manual node selection. Extending this, VideoTree\cite{wang2024videotree} clusters frames based on visual elements and query relevance, advancing interactive narrative capabilities.

Building on these foundations, our system, StoryNavi, emphasises narrative-driven non-linear viewing by exploring how users perceive narratives when interacting with segmented content. By combining segment retrieval with narrative construction, our approach enhances engagement thanks to a cohesive yet flexible video-watching experience.

\section{Prototype User Interface}
\begin{figure*}
    \centering
    \includegraphics[width=\linewidth]{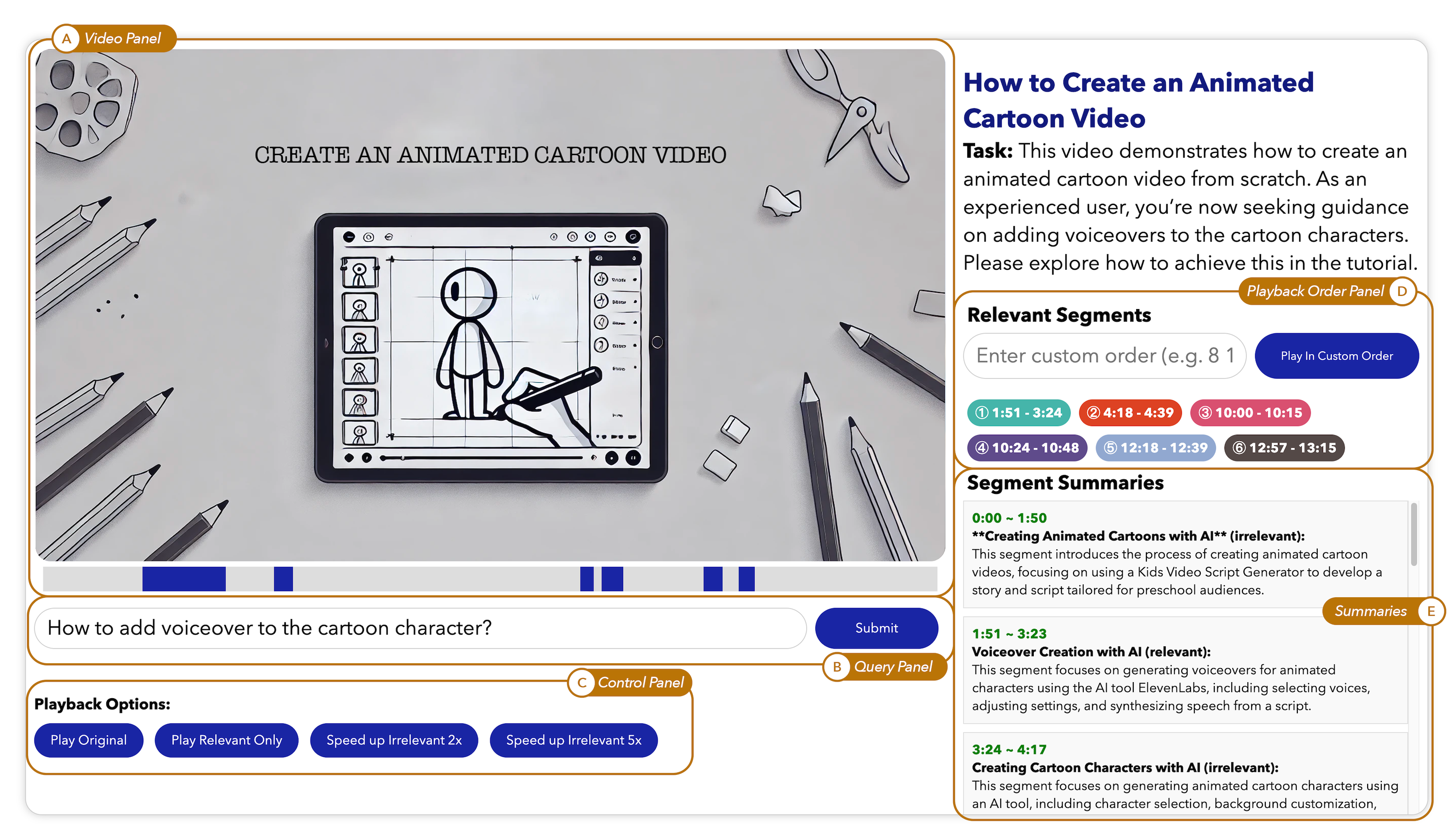}
    \caption{Prototype user interface for query-based video playback. The interface includes multiple panels: (A)Video Panel displaying pre-processed videos with a timeline indicating relevant (blue) and irrelevant (grey) segments, (B)Query Panel where users can input queries to retrieve relevant segments, (C)Control Panel offering playback options for segment-specific control, (D)Playback Order Panel listing all relevant segments with customisable playback order, and (E)Summary Panel providing brief descriptions of both relevant and irrelevant segments for quick content overview. Note: the video content has been replaced with an AI-generated image due to copyright concerns.}
    \label{fig:prototype_demo}
    \Description{Prototype user interface for query-based video playback. The interface includes multiple panels: (A)Video Panel displaying pre-processed videos with a timeline indicating relevant (blue) and irrelevant (grey) segments, (B)Query Panel where users can input queries to retrieve relevant segments, (C)Control Panel offering playback options for segment-specific control, (D)Playback Order Panel listing all relevant segments with customisable playback order, and (E)Summary Panel providing brief descriptions of both relevant and irrelevant segments for quick content overview. Note: the video content has been replaced with an AI-generated image due to copyright concerns.}
\end{figure*}

We developed a web interface that supports text-based queries and features a video playback window. Users can interact with the system by playing, pausing, fast-forwarding, and rewinding the video using their mouse or keyboard as usual. To retrieve relevant video segments, users input queries, which are processed through GPT-4o along with a predefined system prompt. The prototype is illustrated in figure \ref{fig:prototype_demo}.

\textbf{Video Panel.}
Pre-processed videos are displayed using a standard web video player. Below the video, an artificial timeline highlights relevant segments in blue and irrelevant ones in grey. Users can click on any segment to play the corresponding part.

\textbf{Query Panel.}
Users can input queries freely to find relevant segments based on their interest.

\textbf{Control Panel.}
Users can control playback by selecting to either play only relevant segments or speed up irrelevant segments by 2x or 5x. As shown in figure \ref{fig:prototype_demo}(C), the panel offers these options:

\begin{enumerate}
    \item Play Relevant Only: Plays relevant segments continuously.
    \item Speed Up Irrelevant 2x: Speeds up irrelevant segments to 200\%, while relevant segments play at normal speed.
    \item Speed Up Irrelevant 5x: Speeds up irrelevant segments to 500\%, while relevant segments play at normal speed.
\end{enumerate}

\textbf{Playback Order Panel.}
After retrieval, all relevant segments are listed in the Playback Order Panel, as shown in figure \ref{fig:prototype_demo}(D). The panel color-codes and labels each segment with start and end timestamps. Users can click any label to play the segment, which will stop when finished. Segments are numbered, and users can reorder them to customise playback order.

\textbf{Summary Panel.}
Along with the relevant segment labels, summaries of both relevant and irrelevant segments appear in the Summary Panel, as shown in figure \ref{fig:prototype_demo}(E). These brief descriptions help users quickly grasp the content of each section, making it easier to identify the most pertinent segments without having to watch the entire video.

\subsection{Backend Computational Pipeline}
This section outlines the backend processing and pipeline enabling the interactive components described earlier, comprising three key stages: 1) video annotation, 2) segment retrieval, and 3) segment summarisation. OpenAI's GPT-4o model is utilised for all LLM-related tasks discussed herein.

\begin{figure*}
    \centering
    \includegraphics[width=\linewidth]{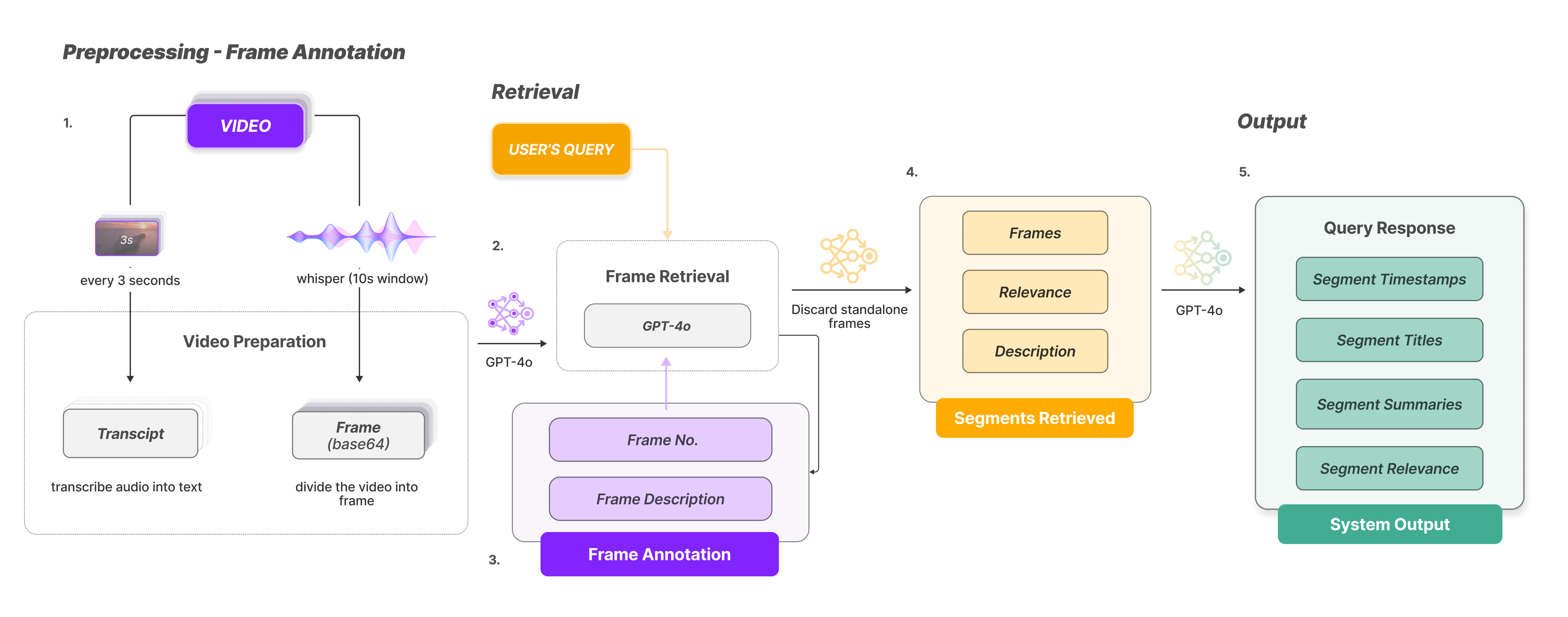}
    \caption{Prototype pipeline. 1) Image and audio extraction from the video. 2) Frame annotations using GPT-4o. 3) Frame Retrieval based on user query. 4) Refine segments. 5) Output results.}
    \label{fig:prototype_pipeline}
    \Description{Prototype pipeline. 1) Image and audio extraction from the video. 2) Frame annotations using GPT-4o. 3) Frame Retrieval based on user query. 4) Refine segments. 5) Output results.}
\end{figure*}

\textbf{Multimodal Video Annotation. }
We initially annotated the frames based solely on visual content, but retrieval accuracy was far from ideal, particularly when causal relationships were not captured or when the visual content remained static for an extended period, while key information was conveyed through the voiceover. Therefore, we applied a multimodal approach. The pipeline begins by capturing a video frame every 3 seconds, encoding each as base64. A 10-second transcript window, spanning 5 seconds before and after each frame, is processed using the Whisper model\cite{radford2023robust}. This method leverages verbal context to enhance user comprehension, with a 10-second window generally sufficient for a complete sentence. The frame and transcript are then fed into the LLM, which generates a concise frame description (up to 50 words) designed to fit within the model's 128K input context window (roughly enough for any video within 90 minutes). Initially, annotations were embedded for future retrieval; however, we opted to retain the original texts due to superior retrieval performance in our preliminary tests.

\textbf{Relevant Segment Retrieval. }Upon receiving a user query, the LLM identifies relevant frames based on the entire annotation file and the query. Although the GPT-4o input context window was sufficient for the entire video annotation database, the frames were retrieved in batches. This approach aligns with recent research showing that LLM comprehension declines significantly as the text length increases\cite{li2023how,li202emodels}.

To ensure a smoother viewing experience, isolated relevant frames with no neighboring relevant frames are discarded. Additionally, if a single irrelevant frame is found between two relevant segments, it is merged into the relevant segments, thereby discarding standalone 3-second (1-frame) relevant segments and merging isolated irrelevant segments into adjacent relevant ones.

\textbf{Segment summarisation. }After identifying the relevant segments, the LLM generates a title and a concise description (up to 40 words) for each segment, including both relevant and irrelevant ones.

\section{Study I - Prototype Iterative Design}
Our initial study aimed to observe and identify interaction design needs between LLMs and users to better support a non-linear viewing experience. Focusing on the interaction model during query-based video watching, this study did not directly contribute to LLM development but rather highlighted the requirements and design guidelines for effective video segment retrieval based on user queries. We began our investigation with an online survey to understand user video-watching habits and preferences, particularly in genres where skipping frequently occurs.

\subsection{Video Watching Survey}
To gain a comprehensive understanding of user video-watching behaviors, particularly in relation to non-adjacent segments, we conducted an online survey with 25 respondents, using a 5-point Likert scale for all rating questions. Initially, we identified the most-watched video genres among participants. The top three genres were: 1) entertainment, 2) vlogs, and food/cooking, and 3) educational tutorials. These findings reflect general viewing preferences.

\textbf{Information Seeking. }Given our focus on non-adjacent segment viewing, we asked participants which video genres they are most likely to seek specific information from. Educational videos ranked highest (mean = 3.909), followed by product reviews (mean = 3.727) and cooking videos (mean = 3.095), indicating that these genres often prompt users to look for particular information rather than passively watch the entire video. Additionally, when asked which genres they are most likely to skip, participants identified news reports (mean = 3.500), product reviews (mean = 3.364), and sports (mean = 3.350) as the top three. The overlap between genres where users seek information and those they skip suggests that the need for targeted retrieval tools is especially relevant in these categories, therefore a guidance for our system scenarios.

\textbf{Seeking-caused Events. }We further explored the motivations behind skipping behaviour, finding that the most common reasons were: 1) Skipping uninteresting parts (N = 20), 2) Already knowing some of the information (N = 17), and 3) Searching for specific answers or content (N = 16). To efficiently locate intended content in videos longer than 15 minutes, users reported employing several strategies. The most frequent method was speeding up playback and waiting for key content (N = 15), followed closely by dragging the play bar at increased speed (N = 14) and using thumbnail previews to skip to relevant sections (N = 14). A subset of users resorted to more arbitrary techniques, such as randomly skipping and looping based on assumptions (N = 9), while others preferred skipping back and forth between segments (N = 5). When asked about the frequency of skipping between multiple segments, participants generally indicated infrequent use of this approach (mean = 2.875).

The results of this survey highlight a clear demand for efficient segment retrieval in video watching, particularly for educational content, product reviews, cooking videos, and may extent to news reports as well as sports videos. Users often engage in active seeking behaviours, such as speeding up playback and dragging the play bar, to bypass irrelevant content and locate specific information. The frequent occurrence of these behaviours underscores the need for more advanced video navigation tools that can better support users in accessing relevant segments without excessive manual intervention. The infrequent use of skipping between multiple segments suggests that current video interfaces may not effectively support this type of navigation, further emphasising the potential value of LLM-driven segment retrieval systems. Future design guidelines should focus on enhancing user control over video content, enabling more seamless access to the information users seek, and reducing the cognitive load associated with navigating lengthy videos.

\subsection{Participants}
To elucidate the advantages and disadvantages of LLM in segment retrieval, we conducted a formative study involving six participants with an age range of 24 to 30 (mean = 27.167, std = 2.640), comprising four males and two females. All participants reported regular daily video-watching habits.

\subsection{Procedure}
We included three distinct video genres—news reports, tutorials, and product reviews—into our study, detailed in table \ref{tab:s1_vid}. For each video, we specified the content that the user should retrieve or emphasise, as if it were their own interests. Participants were tasked with viewing each video twice: initially using YouTube to leverage its native functionalities, and subsequently with our prototype system, allowing a comparative analysis of user behaviours and needs across platforms.

\begin{table*}[]
\begin{tabular}{c|cccll}
\textbf{Video ID} & \textbf{Video Type} & \textbf{Video Title (Duration)}                          & \textbf{URL} &  &  \\ \cline{1-4}
V1-1              & News Report         & How Paris Pulled Off One Of The Cheapest Olympics (744s) & \cite{paris2024}            &  &  \\ \cline{1-4}
V1-2              & Tutorial            & How to Make an Animated Cartoon Video Using Al (999s)    & \cite{cartoon2024}            &  &  \\ \cline{1-4}
V1-3              & Product Review      & New Tesla Model 3 2024 review (1416s)                    & \cite{tesla2023}            &  & 
\end{tabular}
\caption{Three videos used in Study 1}
\Description{Three videos used in Study 1.}
\label{tab:s1_vid}
\end{table*}

\subsection{Results}
Participants provided generally positive feedback on the system's effectiveness in identifying target segments compared to YouTube (\textit{M} = 5.444, \textit{SD} = 0.984), as well as on the overall viewing experience (\textit{M} = 5.500, \textit{SD} = 0.786). The system’s retrieval performance was rated highest for news report videos (\textit{M} = 5.833, \textit{SD} = 0.753), while tutorial videos received the most favorable ratings for overall viewing experience (\textit{M} = 5.667, \textit{SD} = 1.033). These results highlight the system’s potential for non-linear video navigation, particularly in scenarios requiring precise content retrieval.

\subsubsection{YouTube Observations}
Participants primarily used YouTube’s chapters and manual navigation tools (e.g., arrow keys, timeline clicks) to locate content. Speeding up playback (1.75x to 2x) was common when skipping irrelevant sections, particularly in longer videos. However, the predefined chapter titles were often too general, leading participants to engage in extensive manual searching to find specific content. We developed a Chrome extension to collect skipping events during YouTube browsing sessions. Skips shorter than 1 second were discarded, as they were mostly due to continuous right-arrow key presses. We also excluded skips with time gaps longer than 12 seconds from the previous data point, as these were unlikely to be part of an active seeking process. Our analysis shows the average skipping event time gap is \(t = 5.465\) seconds, capped at the 75th percentile.

\subsubsection{Prototype Strengths} We summarise the key findings from testing our prototype, focusing on two main areas: the effectiveness of segment retrieval and the clarity of segment presentation. These findings primarily reflect subjective user feedback, offering insights into how well the system retrieves relevant video segments based on queries and how effectively the segments are presented to enhance the viewing experience as well as content comprehension.

\textbf{Segment Identification and Summaries. }  
Participants found the system's segment identification generally effective. Summaries were particularly appreciated for providing quick overviews of relevant segments, reducing the need for manual searches. However, broken sentences at segment boundaries and incomplete segments were frequent issues, mentioned by five of six participants. As P1 noted, "\textit{I found it disruptive and difficult to follow what was happening during the first few seconds after skipping to a new segment}," with P3 adding, "\textit{Sometimes the subject is missing in the first sentence, which made it unclear what it was referring to}." This often required manual forward/backward skipping to retrieve missing information, disrupting the non-linear viewing experience. 

Three participants requested a general summary of all relevant segments, with P4 stating, "\textit{I would like a brief understanding of the bigger picture before I dive into the details, especially when I don't heavily rely on visual content like in news reports}." They (P1, P4, P6) also emphasised the need for more concise or adaptive segment summaries, particularly in tutorial or educational videos where quick comprehension is crucial. Additionally, two participants noted that irrelevant segments helped cross-check information, enhancing their trust in the system’s retrieval accuracy. 

\textbf{Query Prompting. }  
Most participants preferred using short, keyword-based queries but faced challenges in crafting effective prompts. Four participants revised their queries after initial results, indicating a need for more intuitive query refinement. One concern raised by P3 was being "lazy to type," a sentiment echoed by P5, who stated, "\textit{I'm a lazy person and I don't like typing a long, complete question with explanations just to make sure AI understands me, as the AI should adapt to my habits, not the other way around}." Auto-completion and keyword suggestions were proposed by P1, P3, and P5 as ways to help users formulate more effective queries.

\textbf{Playback and Transitions. }  
Participants expressed strong preference for a smoother transitions between relevant segments, with clearer indications of upcoming skips. Five participants reported that the lack of notifications about these transitions caused confusion. Additionally, all participants found that manually rearranging the playback order of relevant segments was unnecessary and difficult to use, especially for sequential or instructional videos, where maintaining the original order is crucial for comprehension. However, P3 suggested the possibility of rearranging the playback order based on relevance to the query, stating, "\textit{If the segments were played in order of relevance and transitioned smoothly, I might actually prefer that}."

\textbf{Usage Scenarios. }
The system was reported to be potentially useful for structured content like tutorials and educational videos, where participants needed to retrieve specific segments or forgotten content. It also performed well in product reviews, where detailed, segment-based retrieval was essential. Most participants indicated they wouldn't use the system for entertainment purposes. However, P2 mentioned a potential use case for fuzzy queries, stating, "\textit{I would like to find some highlight segments when I don't really remember the details in super long movies, like The Lord of the Rings}." However, the system was less effective for videos with mixed or complicated content, such as news reports as they require a more high-level understanding of the topic.

\subsection{Design Guidelines}
Building on aforementioned user feedback and observations, we propose three key design guidelines to enhance non-linear video navigation and the overall user experience:

\textbf{D1. Provide smooth transitions during skipping events.}
Participants frequently highlighted issues with broken sentences and incomplete segments, which disrupted the viewing experience. Seamless transitions between video segments, especially for those separated by short intervals (less than 6-9 seconds), are essential to maintain narrative coherence. Clear signals before transitions will help users prepare for upcoming skips and ensure continuous playback.

\textbf{D2. Enhance contextual coherence of discrete segments.}
Users struggled with constructing prompts and found it challenging to navigate disjointed segments. To ensure non-linear navigation remains effective, segments should be reorganised based on user queries to form a cohesive narrative. Providing interactive tools like context-aware prompts and keyword suggestions can improve usability, while maintaining continuity across segmented playback enhances understanding.

\textbf{D3. Enable adaptive information presentation.}
Participants valued flexibility in how segment information is presented, especially in tutorial or educational videos. Offering concise, chapter-like titles alongside detailed summaries allows faster navigation and quick reference. An overall summary of relevant segments should also be included, giving users more control over the level of detail they need based on the viewing context.

\section{Backend Pipeline}
With these design guidelines in mind, we updated our backend retrieval pipeline by integrating transcript data not only for frame annotation but also for frame retrieval. Additionally, the retrieved segments are now extended to align with the nearest full sentence at the start and end.

\subsection{Segment Retrieval and Refinement}
\subsubsection{Frame and Transcript Integration.}
During initial testing in Study 1, the frame retrieval prompt yielded suboptimal results, particularly for video genres like news reports, where key information is conveyed primarily through the voiceover rather than visual content. The reliance on visual annotations alone led to the omission of critical details. To address this, we revised our method by integrating transcript data directly into the retrieval prompt, aligning with \textbf{D2}. By including voiceover content within a 5-second window before and after each frame, we ensured that both visual and auditory cues contributed to the segment retrieval process, improving coherence and completeness. This adjustment allowed the system to better capture key information, especially in content-heavy videos like news reports. The revised prompt is detailed in Appendix A.1.

\subsubsection{Voiceover Sentence Completion.}
Study 1 participants consistently emphasised the need for complete sentences in the voiceover when playing back segments, underscoring the importance of narrative flow. The original segment timestamps, based on 3-second intervals, often split sentences, creating a disjointed viewing experience. To ensure a smoother transition and narrative integrity, we revised the segment boundaries to include full sentences, following \textbf{D1}. Using Whisper-1 for transcription, we identified complete sentences (indicated by punctuation) and adjusted the start and end times of segments to align with sentence boundaries. We illustrate the updated process of segment retrieval in figure \ref{fig:mode}.

\subsection{Pipeline Evaluation}
We evaluated our LLM-based retrieval pipeline using 50 queries across 10 videos processed with GPT-4o. Each query's frame selection was repeated five times. The videos, chosen from four genres (Educational, Cooking, Product Reviews, and News Reports) based on user preferences, are detailed in Table \ref{tab:pipeline_vid}. All videos were under 30 minutes, and for each, five queries of varying complexity were created: 2 queries referred to fewer than 3 segments, and 3 referred to more than 3 segments. The ground truth for each query was manually annotated. We calculated recall and precision to assess the pipeline's accuracy in identifying relevant segments:

\begin{equation} \label{recall}
recall = \frac{\text{Correct relevant segments retrieved}}{\text{Total relevant segments}}
\end{equation}

\begin{equation} \label{precision}
precision = \frac{\text{Correct relevant segments retrieved}}{\text{Total segments retrieved}}
\end{equation}

Recall measures the extent of overlap between predicted and relevant segments, while precision reflects how accurately the retrieved segments match the relevant ones.

\begin{table*}
\centering
\begin{tblr}{
  cells = {c},
  cell{2}{2} = {r=2}{},
  cell{4}{2} = {r=2}{},
  cell{6}{2} = {r=4}{},
  cell{10}{2} = {r=2}{},
  hline{2,4,6,10} = {-}{},
}
\textbf{Video ID} & \textbf{Genre} & \textbf{Duration (seconds)} & \textbf{Content Type} & \textbf{URL} \\
V2-1              & Educational    & 2766                        & Sequential            &\cite{pc2021}              \\
V2-2              &                & 999                         & Sequential            &\cite{cartoon2024}              \\
V2-3              & Cooking        & 940                         & Sequential            &\cite{soup2022}              \\
V2-4              &                & 1062                        & Conceptual            &\cite{steak2023}              \\
V2-5              & News Reports   & 626                         & Sequential            &\cite{football2021}              \\
V2-6              &                & 789                         & Conceptual            &\cite{trump2024}              \\
V2-7              &                & 712                         & Conceptual            &\cite{tech2024}              \\
V2-8              &                & 744                         & Conceptual            &\cite{paris2024}              \\
V2-9              & Product Review & 1567                        & Conceptual            &\cite{earbud2024}              \\
V2-10             &                & 1416                        & Conceptual            &\cite{tesla2023}              
\end{tblr}
\caption{Brief description on videos used in pipeline evaluation.}
\label{tab:pipeline_vid}
\Description{Brief description on videos used in pipeline evaluation.}
\end{table*}

\subsection{Retrieval Results}
The pipeline showed reasonable performance with an average recall of 0.886 and precision of 0.682 across the best experiment groups. The standard deviations for recall and precision were 0.042 and 0.065, respectively. The results, detailed in table \ref{tab:video_genre} and table \ref{tab:video_pl_others}, provide insight into the system's performance across various genres, content types, and query types.

\begin{table*}[htbp]
\centering
\resizebox{\textwidth}{!}{%
\begin{tblr}{
  cells = {c},
  cell{1}{1} = {r=2}{},
  cell{1}{2} = {r=2}{},
  cell{1}{3} = {c=4}{},
  vline{2-3} = {1-7}{},
  hline{3} = {-}{},
}
\textbf{Criteria }                     & \textbf{Overall } & \textbf{By Video Genre} &                  &                       &                         \\
                                       &                   & \textbf{Educational}    & \textbf{Cooking} & \textbf{News Reports} & \textbf{Product Review} \\
Observations                           & 50                & 10                      & 10               & 20                    & 10                      \\
Best Recall                            & 0.886             & \textbf{0.940}                   & 0.911            & 0.847                 & 0.885                   \\
Best Precision                         & 0.682             & 0.687                   & 0.635            & \textbf{0.698}                 & 0.691                   \\
Avg Recall Std                         & 0.042             & \textbf{0.024}                   & 0.056            & 0.032                 & 0.064                   \\
Avg Precision Std                      & 0.065             & 0.101                   & 0.075            & \textbf{0.047}                 & 0.056                    
\end{tblr}%
}
\caption{Recall, precision, and standard deviations across different video genres. Recall and precision values represent the best results out of five experimental rounds, while the standard deviations are averaged over the five rounds.}
\label{tab:video_genre}
\Description{Recall, precision, and standard deviations across different video genres. Recall and precision values represent the best results out of five experimental rounds, while the standard deviations are averaged over the five rounds.}
\end{table*}

\begin{table*}[htbp]
\centering
\resizebox{\textwidth}{!}{%
\begin{tblr}{
  cells = {c},
  cell{1}{1} = {r=2}{},
  cell{1}{2} = {r=2}{},
  cell{1}{3} = {c=2}{},
  cell{1}{5} = {c=2}{},
  cell{1}{7} = {c=3}{},
  vline{2-3,5,7} = {1-7}{},
  hline{3} = {-}{},
}
\textbf{Criteria }                     & \textbf{Overall } & \textbf{By Video Content} &                     & \textbf{By Query Type} &                     & \textbf{Segment Number} &                &                          \\
                                       &                   & \textbf{Conceptual}       & \textbf{Sequential} & \textbf{Conceptual}    & \textbf{Procedural} & \textbf{\textless{} 3}               & \textbf{[3,5]} & \textbf{\textgreater{}5} \\
Observations                           & 50                & 30                        & 20                  & 33                     & 17                  & 16                                   & 26             & 8                        \\
Best Recall                            & 0.886             & 0.852                     & \textbf{0.937}               & \textbf{0.891}                  & 0.875               & \textbf{0.933}                                & 0.885          & 0.794                    \\
Best Precision                         & 0.682             & 0.679                     & \textbf{0.686}               & 0.670                  & \textbf{0.705}               & \textbf{0.732}                                & 0.701          & 0.519                    \\
Avg Recall Std                         & 0.042             & 0.051                     & \textbf{0.028}               & 0.044                  & \textbf{0.038}               & \textbf{0.029}                                & 0.040          & 0.075                    \\
Avg Precision Std                      & 0.065             & \textbf{0.057}                     & 0.078               & \textbf{0.060}                  & 0.077               & 0.082                                & 0.062          & \textbf{0.032}                    
\end{tblr}%
}
\caption{Recall, precision, and standard deviations by video content type, query type, and segment number. Recall and precision values represent the best results out of five experimental rounds, while the standard deviations are averaged over the five rounds.}
\label{tab:video_pl_others}
\Description{Recall, precision, and standard deviations by video content type, query type, and segment number. Recall and precision values represent the best results out of five experimental rounds, while the standard deviations are averaged over the five rounds.}
\end{table*}

\textbf{Video Genre Comparison. }Among the four genres, Educational videos achieved the highest recall, but ANOVA tests showed no significant differences across genres for recall (F = 1.37, p = 0.27), precision (F = 0.25, p = 0.86), or standard deviations of recall and precision.

\textbf{Video Content Type. }We categorised the video content into two types: conceptual and sequential. Conceptual videos focus on explaining ideas or abstract concepts, while sequential videos present information in a step-by-step manner, often following a logical or chronological order. A t-test indicated a significant difference in recall between these two content types (t = 2.01, p = 0.01), but no significant differences in precision (t = 2.01, p = 0.90). This suggests that the pipeline is more effective at retrieving all segments relevant to user interests in sequential videos than in conceptual ones. Additionally, the t-test showed no significant differences in the average standard deviation of recall (t = 2.01, p = 0.11) or precision (t = 2.04, p = 0.35), indicating that the stability of LLM retrieval is not sensitive to video content type.

\textbf{Query Type. }We also classified user questions into two types: procedural and conceptual. Procedural questions typically involve specific procedures or actions ("how-to" questions), while conceptual questions require explanations of ideas, methodologies, or strategies. The t-test presented no significant differences between these two question types in all four criteria: recall of the best group out of 5 experiments (t = 2.05, p = 0.69), precision of the best group out of 5 experiments (t = 2.06, p = 0.57), average standard deviation of recall across 5 experiments (t = 2.02, p = 0.70), and average standard deviation of precision across 5 experiments (t = 2.08, p = 0.51).

\textbf{Ground Truth Segment Numbers. }We further examined the impact of the number of segments in the video corresponding to each question on the LLM retrieval results. We divided segment numbers into three groups: fewer than 3 segments, 3 to 5 segments, and more than 5 segments. Significant differences were found across the three groups in both recall (F = 3.39, p = 0.04) and precision (F = 4.11, p = 0.02). As the segment count increased, both recall and precision decreased. However, there were no significant differences in the average standard deviation of recall (F = 2.20, p = 0.12) or precision (F = 1.28, p = 0.29), suggesting that the stability of LLM retrieval is not dependent on segment numbers.

\textbf{Accuracy.} A key finding was that recall consistently outperformed precision across both overall and group-specific results. Retrieved segments often included content beyond what the user intended to watch, which contributed to lower precision. This was partly due to post-processing steps like dropping standalone frames, merging close segments, and completing broken sentences. Upon further analysis, we found that LLMs sometimes prioritised specific keywords from the query over the overall context, leading to irrelevant frame retrieval. We attempted to address this by optimising the retrieval prompt through a chain-of-thought approach\cite{chainthought,suzgun2022chal}, where the model first analysed what constitutes a satisfactory answer and then selected frames accordingly. However, this approach resulted in even lower recall and precision.

We also tested a high level top-down approach, generating a high-level narrative for the entire video, and then respond to the user query solely on the whole video narrative, similar to human reasoning, and retrieving frames with the this responded narrative. However, this led to lower retrieval accuracy compared to our current bottom-up method. We attribute this to LLMs' difficulty in balancing long-context understanding with fine-grained detail preservation. This challenge is particularly evident when maintaining coherence across extended video segments while accurately identifying frames, as supported by recent studies \cite{liu2023los,li2023how}.

These issues can be attributed to 1) the inherent limitations of current LLMs in processing long contexts and complex information, and 2) loss of key details during the abstraction process, leading to imprecise retrieval. Additionally, since all questions and ground truths were manually annotated, we recognise that the subjective nature of the ground truth may vary based on individual interpretations.

\begin{figure*}
    \centering
    \includegraphics[width=\linewidth]{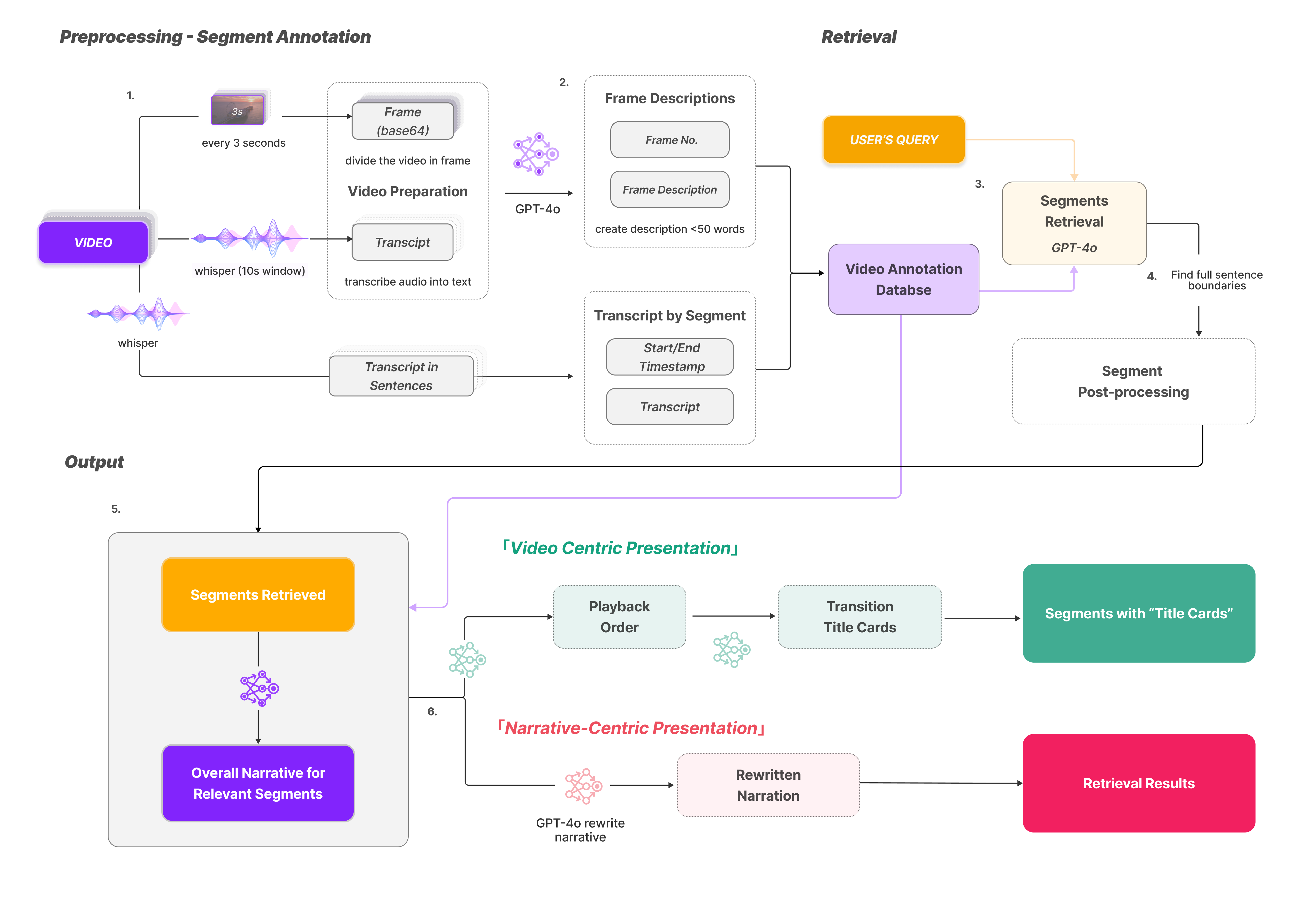}
    \caption{Illustration of StoryNavi pipeline. 1) Image and audio extraction from the video. 2) Frame Annotation using GPT-4o and transcription using Whisper. 3) Frame retrieval based on user query. 4) Refine segments. 5) Narrative generation. 6) Playback Mode, either video-centric or narrative-centric}
    \label{fig:system_pipeline}
    \Description{Illustration of StoryNavi pipeline. 1) Image and audio extraction from the video. 2) Frame Annotation using GPT-4o and transcription using Whisper. 3) Frame retrieval based on user query. 4) Refine segments. 5) Narrative generation. 6) Playback Mode, either video-centric or narrative-centric}
\end{figure*}

\section{StoryNavi}
StoryNavi (figure \ref{fig:teaser}) is a query-based video interface designed to support non-linear video viewing experiences, enabled by the updated backend implementation. After retrieving relevant segments based on user queries, the system generates an overall narrative to ensure a cohesive viewing experience. To examine how narrative structures enhance non-linear viewing, we propose two modes of segment Playback: 1) \textbf{Video-centric}, which preserves the original audio while incorporating artificial transitions using "title cards" for smooth continuity between segments, and 2) \textbf{Narrative-centric}, where the original audio is replaced by a speech synthesiser, with a stronger emphasis on the narrative flow across segments.

\subsection{Narrative-driven Playback Order}
Participants in Study 1 expressed dissatisfaction with the manual manipulation of playback order but highlighted the need for a cohesive understanding of discrete segments. In response, we adopted a narrative-driven approach, rearranging relevant segments to create a more logical and unified viewing experience. Specifically, LLM-generated narratives, tailored to user queries, ensure smooth semantic transitions between segments. This narrative-driven structure is designed to maintain the viewer's focus and comprehension, particularly addressing the concerns outlined in \textbf{D2}. This approach applies to both the video-centric and narrative-centric playback modes.

\begin{figure*}
    \centering
    \includegraphics[width=0.9\linewidth]{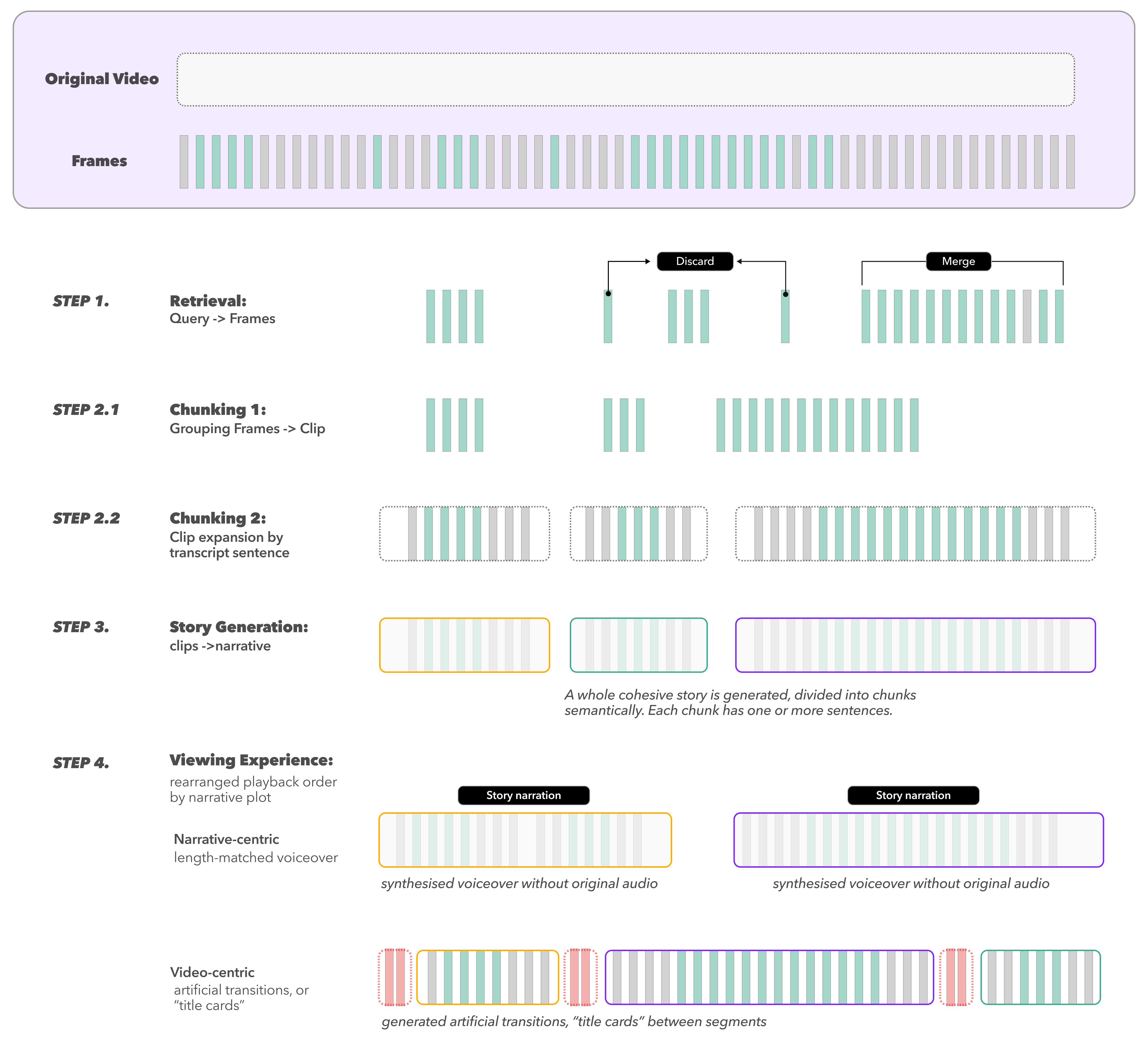}
    \caption{Illustration of segment retrieval and construction of two playback modes.}
    \label{fig:mode}
    \Description{Illustration of segment retrieval and construction of two playback modes.}
\end{figure*}

\subsection{Video-centric Playback}
\begin{figure*}
    \centering
    \includegraphics[width=\linewidth]{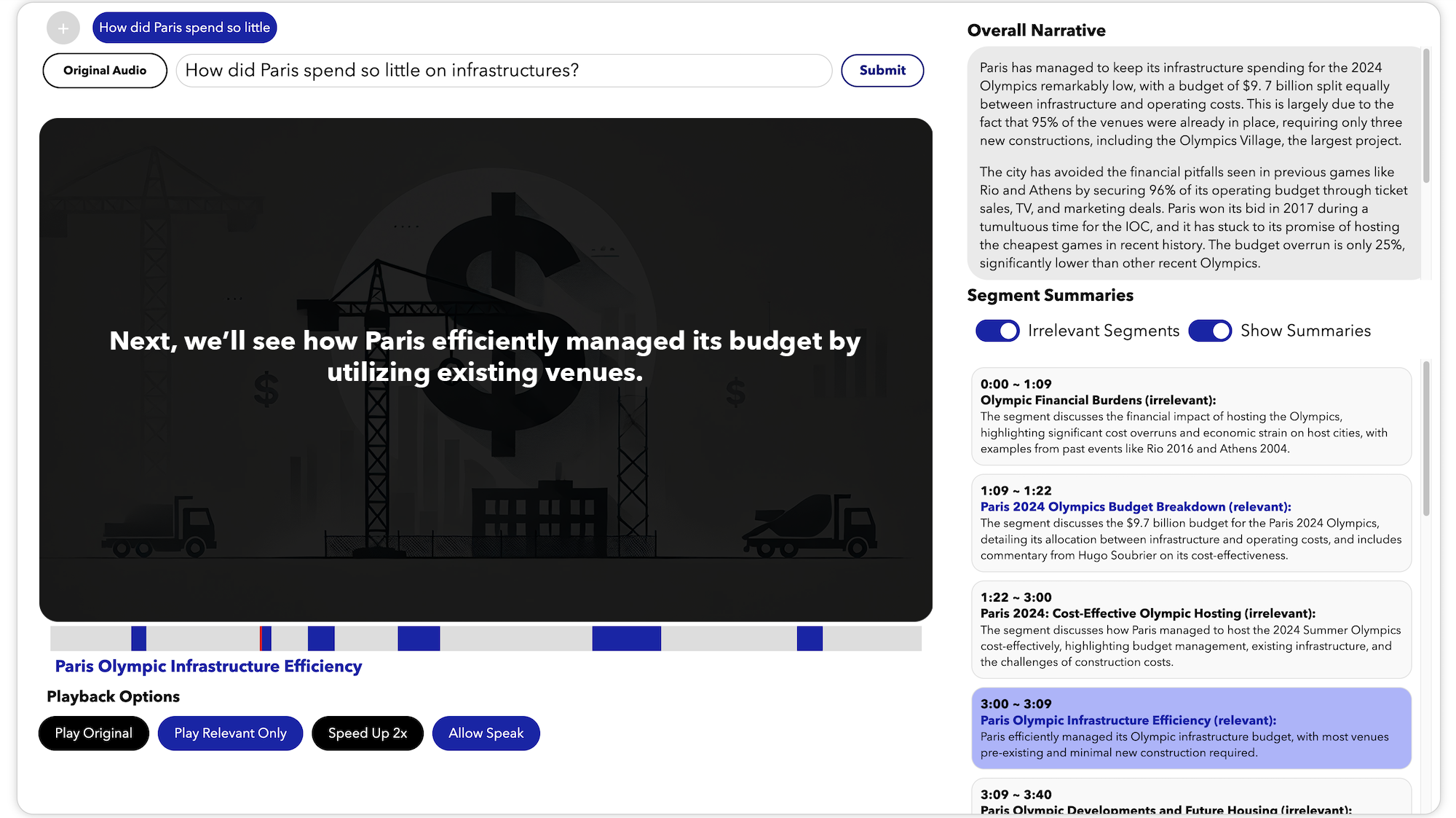}
    \caption{Screenshot of video-centric playback mode, featuring an artificial “title card” between segments that displays LLM-generated transition sentences. Note: the video content has been replaced with an AI-generated image due to copyright concerns.}
    \label{fig:video_centric_demo}
    \Description{Screenshot of video-centric playback mode, featuring an artificial “title card” between segments that displays LLM-generated transition sentences. Note: the video content has been replaced with an AI-generated image due to copyright concerns.}
\end{figure*}

In Study 1, participants emphasised the significance of original audio in various contexts, such as tutorial videos, where the step-by-step instructions are critical. However, they also requested smoother transitions between segments for a more continuous experience \textbf{(D1)}. To address this, we introduced artificial transitions in the form of LLM-generated "title cards." These title cards provide a brief overview of the preceding segment and a preview of the next, helping users set expectations for the upcoming content. By connecting discrete relevant segments into a continuous video, these transitions create a feeling of chapter-based progression. The title card content is generated based on the titles and summaries of consecutive relevant segments, following their playback order, as shown in figure \ref{fig:video_centric_demo}. We also illustrate this process in figure \ref{fig:mode}.

\subsection{Narrative-centric playback}
\begin{figure*}
    \centering
    \includegraphics[width=\linewidth]{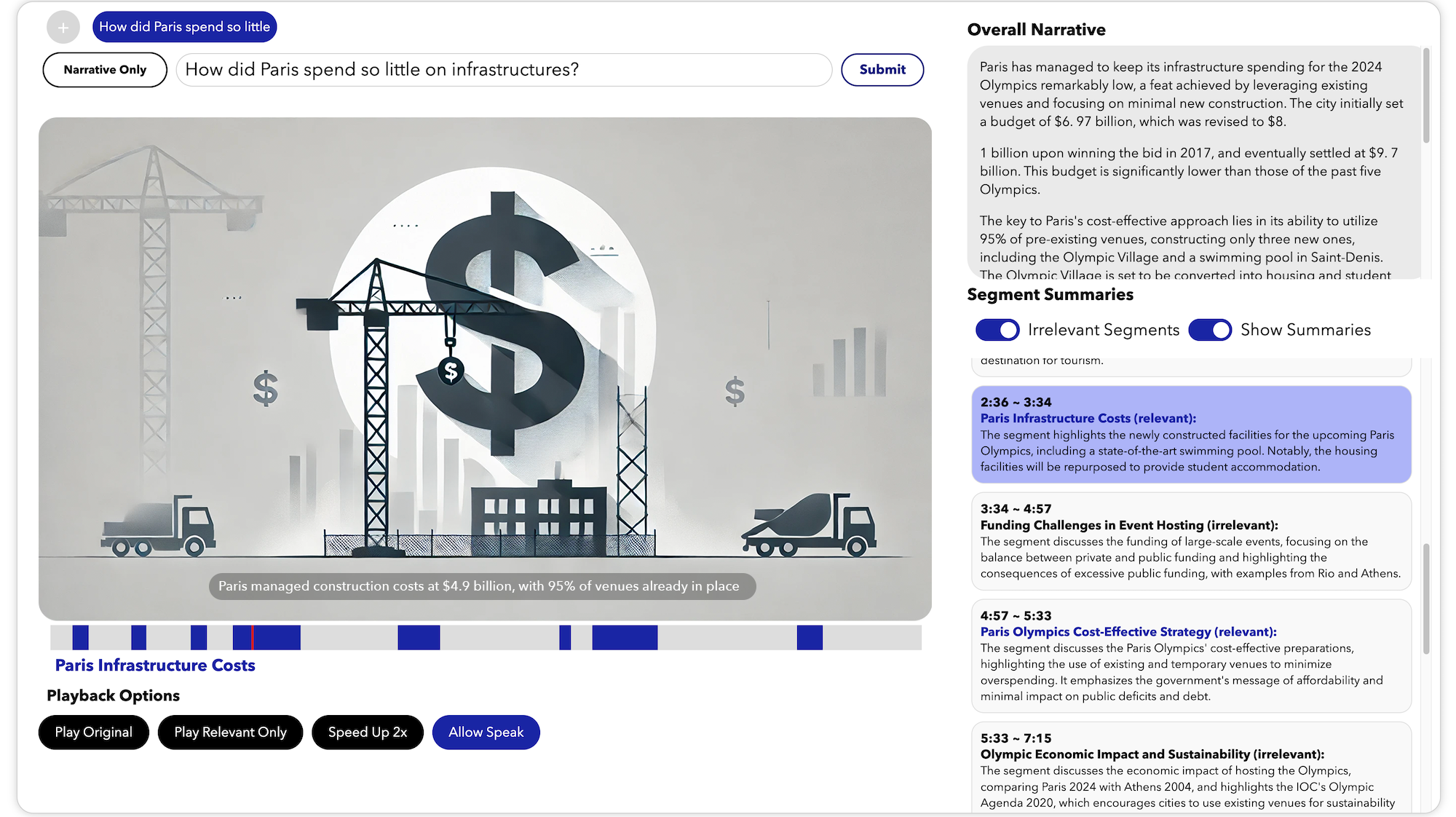}
    \caption{Screenshot of narrative-centric playback mode, with rewritten length-matched synthesised voiceover. Note: the video content has been replaced with an AI-generated image due to copyright concerns.}
    \label{fig:narrative_centric_demo}
    \Description{Screenshot of narrative-centric playback mode, with rewritten length-matched synthesised voiceover. Note: the video content has been replaced with an AI-generated image due to copyright concerns.}
\end{figure*}

For videos containing complex information that requires more than just a visual or text transition, we developed the narrative-centric mode. In this mode, the original audio is replaced with a speech synthesiser, and the segments are rewritten into a cohesive, flowing narrative. The narrative is semantically divided into chunks, with each chunk corresponding to one or more relevant segments identified during retrieval. This structure allows for a more fluid storytelling experience, as the speech synthesiser can span across multiple segments, connecting them through a coherent narrative thread. Unlike the video-centric mode, where segments are only connected by transitions, the narrative-centric mode groups segments based on their semantic content, allowing the narrative to guide the viewing experience. This approach is illustrated in figure \ref{fig:narrative_centric_demo}. We also illustrate this process in figure \ref{fig:mode}.

\section{Study II - User Evaluation} We evaluated the effectiveness of our system through a lab study. Our evaluation aimed to assess the narrative-driven experience offered by StoryNavi, presented in two distinct modes: narrative-centric and video-centric. We focused on measuring 1) the effectiveness of StoryNavi’s query-based interface in reducing user effort and improving comprehension in non-linear video navigation, and 2) how playback mode interacts with different video and query types. The rationale behind this approach was to understand how varying levels of narrative coherence and video fidelity impact user engagement and comprehension. Results are framed against the design guidelines for smooth transitions during skipping \textbf{(D1)} and maintaining contextual coherence across discrete segments \textbf{(D2)}, demonstrating the system’s adherence to these principles.

\subsection{Participants}
We recruited 8 participants (5 males, 3 females, aged 23-30) from the campus, all of whom watch videos daily. Additionally, all participants had prior experience with LLMs, primarily for academic or entertainment purposes. They noted LLMs excel at summarisation but sometimes may behave unpredictably and incorrectly.

\subsection{Study Procedure}
The study lasted between 1.5 and 2 hours per participant. After a 5-10 minute demonstration of StoryNavi, participants followed a counterbalanced within-subject design. Eight videos from four genres were used, with each participant watching one video per genre. For each video, two pre-determined queries were provided—one specific and detailed, the other more conceptually general. The query-relevant segments, along with narratives and summaries, were retrieved in advance to minimise LLM performance bias in this experience evaluation. A free exploration session allowed participants to select any video and generate real-time query results using StoryNavi. To maintain engagement, participants answered simple questions based on the presented segments during each query session. After each session, participants completed a 7-point Likert scale questionnaire covering system usability, task workload, and trust in automation. The study concluded with 15-minute semi-structured interviews to gather detailed feedback on the system's usability and performance.

\section{Results and Findings}
Participants generally found StoryNavi helpful for identifying relevant segments while navigating videos. A key strength of the system was its ability to maintain a continuous narrative during non-linear navigation, which proved especially useful when participants encountered complex or disconnected segments. To compare participant ratings, we first applied the Wilcoxon Signed-Rank test with an alpha level of 0.05. No significant differences were observed between the playback modes across all questions (p > 0.05), suggesting that the impact may be more complex than initially expected.

\subsection{General Overview of User Experience}
\begin{figure}
    \centering
    \includegraphics[width=\linewidth]{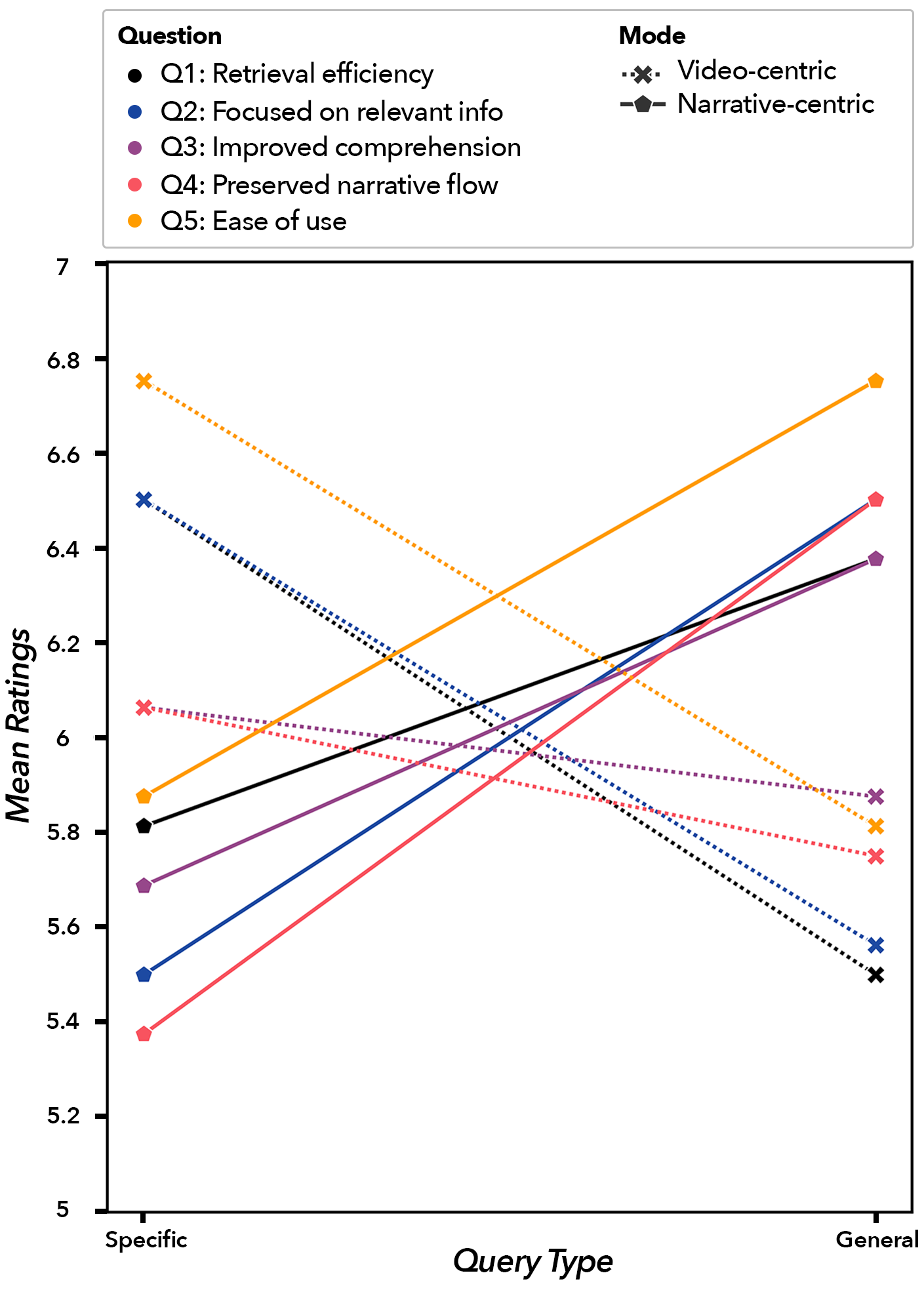 }
    \caption{Interaction plot illustrating the effect of Query Type and Playback Mode on system usability-related questions (Q1-Q5), including efficiency in identifying target information, focus on relevant information, improvement in content comprehension, preservation of narrative flow, and overall ease of use.}
    \label{fig:result1_plot}
    \Description{Interaction plot illustrating the effect of Query Type and Playback Mode on system usability-related questions (Q1-Q5), including efficiency in identifying target information, focus on relevant information, improvement in content comprehension, preservation of narrative flow, and overall ease of use.}
\end{figure}

\begin{figure}
    \centering
    \includegraphics[width=\linewidth]{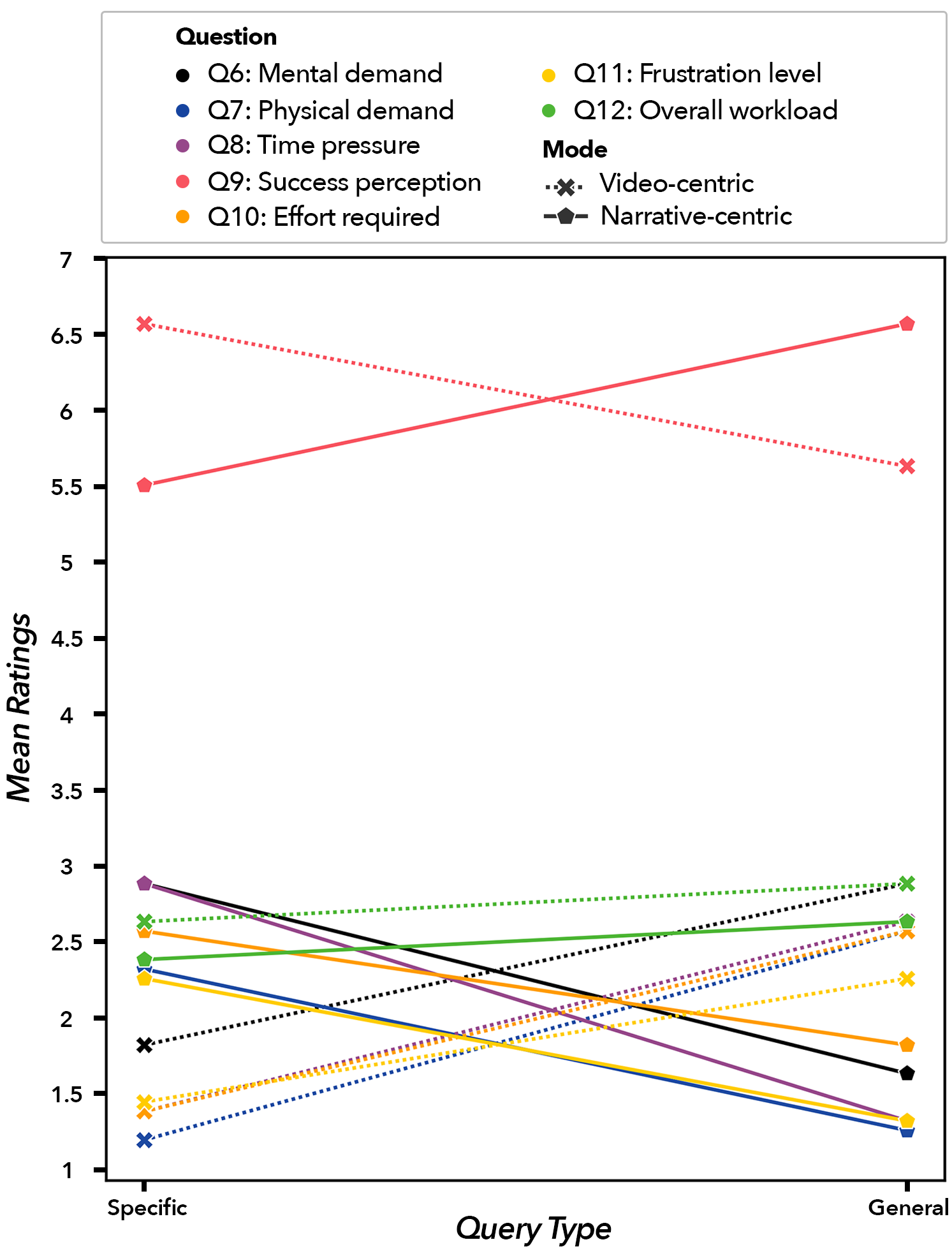 }
    \caption{Interaction plot illustrating the effect of Query Type and Playback Mode on workload-related questions (Q6-Q12), including mental and physical demands, time pressure, perceived success, effort, frustration, and overall workload.}
    \label{fig:result2_plot}
    \Description{Interaction plot illustrating the effect of Query Type and Playback Mode on workload-related questions (Q6-Q12), including mental and physical demands, time pressure, perceived success, effort, frustration, and overall workload.}
\end{figure}

\begin{figure}
    \centering
    \includegraphics[width=\linewidth]{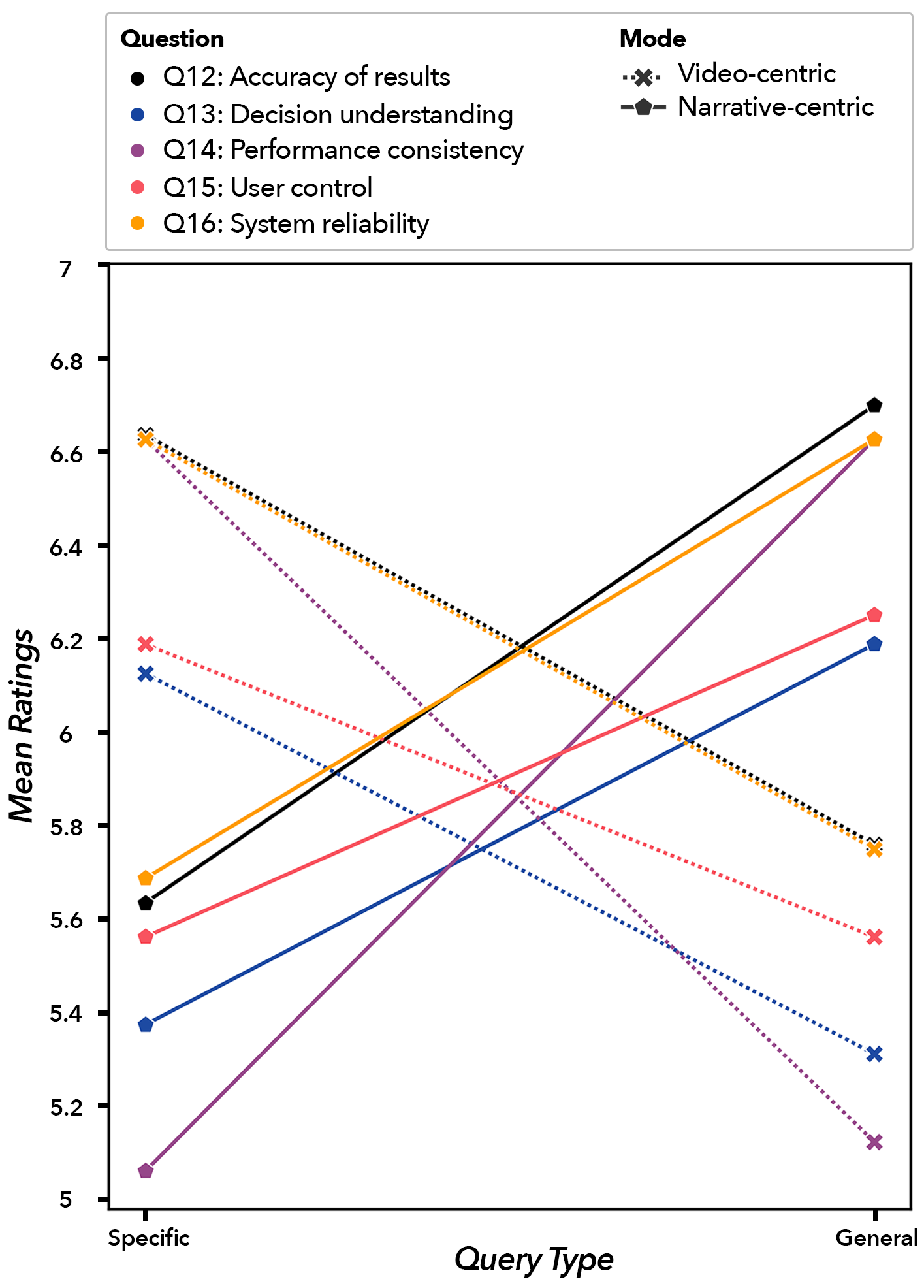 }
    \caption{Interaction plot illustrating the effect of Query Type and Playback Mode on system/AI trust-related questions (Q13-Q17), covering system accuracy, decision understanding, performance consistency, user control, and system reliability.}
    \label{fig:result3_plot}
    \Description{Interaction plot illustrating the effect of Query Type and Playback Mode on system/AI trust-related questions (Q13-Q17), covering system accuracy, decision understanding, performance consistency, user control, and system reliability.}
\end{figure}

To explore the impact of multiple factors on user experience, we performed an Aligned Rank Transform (ART) ANOVA. This analysis allowed us to evaluate the effects of Playback Mode (Video-centric vs. Narrative-centric), Query Type (Specific vs. General), and Video Content on user ratings across 17 questionnaire items. The analysis includes system usability (interaction plot for Q1-Q5 in figure \ref{fig:result1_plot}), task workload (interaction plot for Q6-Q12 in figure \ref{fig:result2_plot}), and trust in automation (interaction plot for Q13-Q17 in figure \ref{fig:result3_plot}). Except for task workload (Q12), all questions revealed an interaction effect between Query Type and Playback Mode. The following subsections present both the qualitative insights and quantitative results in detail.

\subsection{Narrative-Preserving and System Usability}
Both playback modes were rated positively for maintaining narrative continuity, regardless of the actual playback order of segments. As P1 remarked, "\textit{I did not notice that unless I saw the timeline skipped}", added by P4, "\textit{the title cards offered me what to expect so that I won't be lost.}" \textbf{(D1)}. 

In Narrative-centric mode, participants reported improved flow between segments, making it easier to follow disjointed or complex content. P2 commented, "\textit{I feel like watching a complete and cohesive video instead of discrete segments...}," a sentiment echoed by P3, P4, and P6. This preference was reflected in significantly better ratings for Q5 - Ease of Use in the Narrative-centric mode (F(1, 28) = 13.84, p < 0.001), with users noting that they felt "\textit{less eager to look into other information such as segment summaries.}"
This mode also reduced Q7 - Physical Demand (F(1,28) = 4.80, p = 0.037), showing that a coherent narrative aids in navigating non-linear video content. These findings suggest that a well-structured narrative plays a critical role in helping users efficiently navigate non-linear video content, demonstrating StoryNavi's capability to offer a coherent understanding of discrete segments \textbf{(D2)}.

However, five participants pointed out alignment issues between the narrative and visual content, with P3 noting, "\textit{I was sometimes confused when the visual didn't perfectly align with the narration.}" This feedback highlights the importance of temporal coherence between the voiceover and visual elements to enhance user comprehension. Additionally, five participants mentioned that the voice of the synthesiser lacked a natural, human-like quality, which affected their level of engagement. The synthesiser's voice was described as lacking emotion and emphasis on key words. As P2 stated, "\textit{it sounds like simply reading a paragraph without any focus,}" while P3 added, "\textit{it sometimes makes me less focused on the content, and instead I find myself trying hard to understand the synthesiser itself.}" This feedback suggests that improving voice synthesis would further enhance user engagement.

Despite these concerns, participants expressed greater trust in the system’s outputs in the Narrative-centric mode, which received higher ratings for Q12 - Trust in Accuracy of Results (F(1, 28) = 5.41, p = 0.027). This shows that a well-maintained narrative helps users comprehend and trust the system's outputs.

\subsection{Impact of Query Type and Retrieval Accuracy}
Query Type also had a significant influence on user experience. Specific queries resulted in more positive outcomes, as they led to clearer, more focused responses from the system. Participants found the system easier to use with specific queries (F(1, 28) = 13.84, p < 0.001), likely due to the precision and relevance of the retrieved segments. P4 observed, "\textit{When my query roughly aligns with the structure of the original video, the retrieval performance was superior and easier to follow, for example, in the football video introducing GOATs at all positions.}" This clarity translated into higher Q12 - Trust in Accuracy of Results (F(1, 28) = 6.00, p = 0.021), with users feeling more assured that the system was providing the most relevant information for their needs.

However, challenges arose with less precise retrieval, especially when irrelevant information appeared at the beginning or end of segments. As P6 mentioned, "\textit{One segment started with ‘his,’ but I didn't know who it referred to.}" This highlights the need for more accurate segment boundaries to ensure clarity and improve user understanding.

\subsection{Contextual Influence of Video Content}
Video content also influenced participant ratings. Videos that were more structured or conceptually aligned with the system’s narrative flow (e.g., PC assembly, football GOATs) resulted in better comprehension, as reflected in Q13 - Understanding AI Decision-Making (F(7,22.64) = 2.70, p = 0.034) and Q14 - System Performance Consistency (F(7,22.64) = 3.20, p = 0.016).

For more abstract or dense content, such as political interviews, participants preferred the Narrative-centric mode, which presented background and context more clearly. As P2 stated, "\textit{For the Trump interview news video, I can understand the main idea better with the Narrative-centric mode, as the original sentences were not as clear at first.}" P3 added, "\textit{The narrative-centric mode is able to present the background, events, and consequences much clearer.}" This was especially true when General queries were used, which often resulted in broader, less focused segment retrieval, increasing Q6 - Mental Demand (F(1, 4) = 9.28, p = 0.038), as users needed to integrate broader, less directly relevant information.

\subsection{User Preferences and Engagement Based on Video Type}
While participants generally preferred the Narrative-centric mode for complex, disjointed videos, they favoured the Video-centric mode for simpler, instructional videos. Moreover, segment titles and summaries were utilised more frequently in the Video-centric mode. Several participants expressed a preference for quick and concise understanding through the segment summary function.   This was evident from user comments, such as P7’s preference for quickly browsing segment titles rather than viewing summaries: "\textit{When I need to understand complex information, I need some context, even if it's not directly related to my query.}" \textbf{(D3)}

This suggests that while the Narrative-centric mode excels in providing coherence for complex content, the Video-centric mode is more suitable for straightforward instructional material.

\subsection{LLM Performance and Trust in Automation}
While participants were aware of the LLM’s role in generating summaries, there was general trust in its ability to effectively summarise complex information. Five of the eight participants expressed confidence in the LLM’s capabilities, with comments like, "\textit{from my previous experience with LLMs, they are very capable and reliable on summarisation tasks.}" However, three participants (P1, P6, P8) noted that the summaries sometimes lacked emotional nuance or a "human touch," making the content feel somewhat artificial. As P1 remarked, "\textit{it's definitely something an LLM writes.}" These participants suggested that improvements in the distinctiveness and logical flow of the summaries would enhance cohesion and make the system feel more intuitive to follow.

Interestingly, P4 frequently explored irrelevant segments, suggesting that including irrelevant information might help build trust in the system. As P4 explained, "\textit{I just don’t trust it to find all the segments I want, especially for some queries. Every person has a different understanding, for example, well-being/happiness means something very different to everyone.}" This feedback points to the importance of long-term user interaction with the system to build trust and understand its limitations, especially for subjective or abstract queries.

\subsection{Implications for System Usability and Usage Scenarios}
Overall, the system was rated as user-friendly and effective in helping participants locate key information within videos. It was particularly useful in situations where users needed to review specific segments or quickly gather information from long or complex videos. The ability to streamline this process through narrative-preserving mechanisms was highly valued, as it reduced the cognitive load required for non-linear navigation. As P1 described, "\textit{normally I need to manually remember or mark down the key timestamps, now I can quickly locate the segment, even if not perfectly precise, as long as the information I want is included.}" Similarly, P5 noted, "\textit{it somehow looks like automatically generated notes that are closely related to my interests.}"

Participants identified several potential usage scenarios where the system would be especially beneficial, including instructional videos and product reviews, where users often seek specific information. While the system was seen as effective for simplifying information retrieval, improvements in retrieval accuracy could enhance its utility for tasks requiring high precision. Nonetheless, the system was regarded as an efficient tool for simplifying the process of retrieving information from complex videos, particularly when combined with narrative-preserving techniques that ensured a coherent understanding of the content.

\section{Discussion}
This study offers valuable insights into the challenges and opportunities of non-linear video navigation, particularly when integrating narrative-preserving mechanisms. StoryNavi’s approach to maintaining narrative coherence during non-linear navigation demonstrates significant potential to enhance user engagement, comprehension, and trust in video systems, while also highlighting areas for further refinement.

\subsection{The Role of Narrative Continuity in Non-Linear Navigation}
A central finding of this study is the importance of narrative continuity in facilitating non-linear video navigation. Using a cohesive narrative as a guiding framework, especially in discrete content, has notable implications for cognitive load and information retention. Traditional segmented video interfaces often leave users disoriented, requiring them to reconstruct the storyline from discrete pieces. StoryNavi's approach aligns with the "narrative grammar" concept\cite{cohn2017drawing}, which organises sequential content into meaningful stories even when presented out of order, as demonstrated in this study. This structure goes beyond simple linear transitions, enabling users to perceive deeper connections between segments even when they are presented non-chronologically\cite{cohn2017drawing}, as tested in our study. Similarly, MovieDreamer \cite{zhao2024ment} highlights the importance of global narrative coherence in generating long videos that are easier to understand through diffusion-based rendering.

However, misalignment between the synthesised voiceover and visual content did occasionally disrupt the narrative flow. Users reported moments where the synthesised voice did not match the visual content, creating a disconnect. Addressing this through more precise narration processes—possibly integrating frame-specific synchronization between visuals and narration—will be critical in improving the overall coherence of future versions.

\subsection{Balancing Automation, Trust, and User Control}
This study also highlights the balance needed between automation and user control in AI-driven systems like StoryNavi, which uses LLMs for segment retrieval and narrative generation. While automation reduces manual effort and increases efficiency, challenges arise with abstract or subjective queries, as P4 noted concerns about the system’s ability to capture all relevant content. This underscores a limitation of current AI: aligning algorithmic outputs with human interpretation.

Trust in automated systems is built through transparency and predictability. Participants, especially P4 and P5, expressed the need for clearer feedback on how content is retrieved and why certain segments are selected. This aligns with prior research\cite{glikson2020human} on the importance of transparency, reliability, and tangibility in fostering cognitive trust in AI. Future designs should emphasise giving users the ability to refine and adjust results, blending automation with user agency to create a more human-centred system.

\subsection{Emotional Engagement in Human-LLM Interaction}
A recurring theme in participant feedback was the emotional disconnect caused by the synthesised voice in the Narrative-centric mode. Although the technical efficiency of the LLM was praised, the lack of emotional nuance in the synthesised speech detracted from user engagement. This highlights the broader role of affective-computing in human-LLM interaction, where emotional resonance plays a significant part in user satisfaction. Users seek not just functional outputs but a natural, emotionally engaging interaction.

To address this, future systems should integrate more sophisticated text-to-speech technologies that incorporate emotional cues such as tone modulation and emphasis. Research in this area, such as VocaListener and VocaWatcher\cite{goto2012vocalistener}, has demonstrated the potential to imitate human singing and facial expressions. Incorporating such improvements could enhance user engagement and bridge the gap between user expectations and machine-generated outputs.

\section{Limitations and Future Work} 
StoryNavi, is an early attempt at non-linear video navigation, and as such faces certain limitations that present avenues for future research and development.

\subsection{Granularity in LLM Retrieval Pipeline} 
A key limitation of StoryNavi is the granularity of its LLM-powered segment retrieval pipeline. While predefined segment boundaries work well for specific queries, they often include extraneous content when handling abstract or nuanced queries, disrupting narrative coherence. As P3 suggested, adding contextual sentence completion could address this issue. Future work should explore more adaptive models that adjust granularity based on query complexity. Techniques like multi-label topic modeling\cite{multiLabel} or semantic chunking\cite{pradhan2005semantic} could refine segment precision. Offering users manual control over granularity could also improve the balance between detail and context.

\subsection{Synthesised Voice and User Engagement}
The synthesised voice in the Narrative-centric mode lacked emotional nuance, which affected user engagement. Participants found the voice monotonous and robotic, making it difficult to focus on the content. Future iterations should incorporate advanced text-to-speech technologies that enhance emotional expressiveness and offer personalisation options for voice style or tone, making the narrative more natural and engaging.

\subsection{Adaptability to Various Video Types and Queries}
Although the study included 10 videos, future work should expand to cover a wider range of videos and participants, given the diversity of video content. An auto mode recommendation system based on query and video content would improve adaptability, ensuring StoryNavi selects the most suitable mode for each scenario.

\subsection{Unified Timeline} 
\begin{figure}
    \centering
    \includegraphics[width=\linewidth]{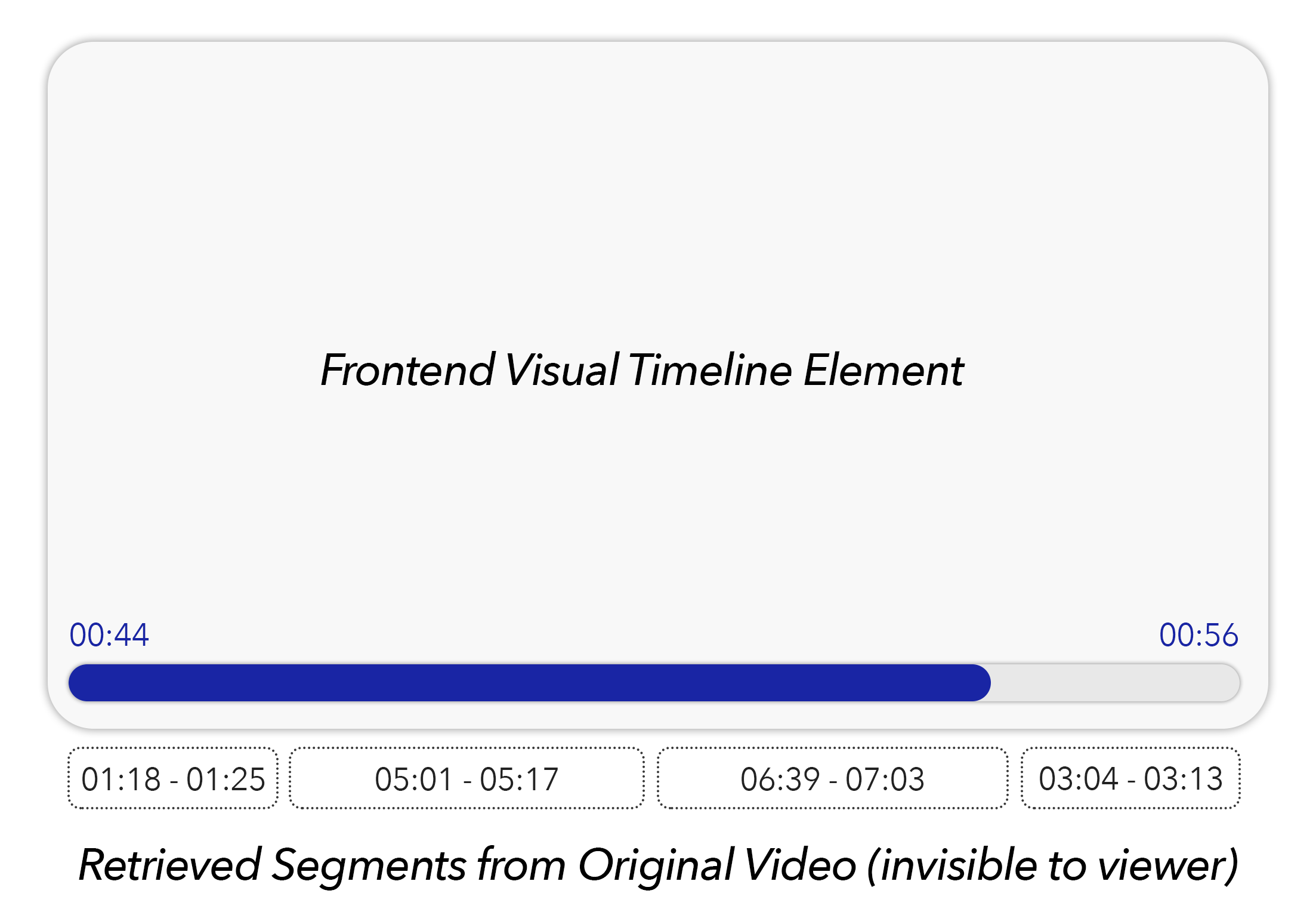}
    \caption{Proposed unified timeline that offers seamless segment transitions.}
    \label{fig:unifiedTL}
    \Description{Proposed unified timeline that offers seamless segment transitions.}
\end{figure}

Participants in the study reported that they did not notice changes in the playback order unless they saw the timestamp jump, which led to confusion during certain moments. For instance, P6 commented, "\textit{I'm not sure if all relevant segments have been played or if some segments are being replayed, and whether this is a bug due to the experimental setup.}" To address this issue and improve narrative coherence, we propose a Unified Timeline design as illustrated in figure \ref{fig:unifiedTL}.

In this approach, relevant segments are no longer marked on the original video timeline. Instead, a new timeline replaces the original, reflecting the narrative flow and displaying segments in the order they are played. This unified timeline allows users to scrub through content as a continuous sequence, independent of the original video’s segment order. By synchronising the visual timeline with the narrative structure, users can navigate more fluidly without being disrupted by timestamp jumps, minimizing confusion and fostering a more cohesive and immersive viewing experience.

\section{Conclusion}
In this paper, we introduced StoryNavi, a narrative-driven, LLM-powered system designed to enhance non-linear video navigation by maintaining narrative coherence across disjointed content. Our findings show that preserving narrative structure significantly improves user comprehension and reduces cognitive load, especially for complex or disjointed videos. While StoryNavi effectively facilitates navigation, challenges remain, such as refining retrieval granularity, improving the quality of the synthesised voice, and enhancing adaptability to diverse video types. By addressing these limitations, StoryNavi can further improve the video-watching experience, making it more intuitive and engaging for users as they search for relevant information within a growing landscape of video content.

\bibliographystyle{ACM-Reference-Format}
\bibliography{mybib}

\appendix

\section{Prompts}

\subsection{Frames Retrieval}
You are given a video frame description every 3 seconds and the voiceover around this frame (the voiceover time range is within 5 seconds before and after each frame), along with the user's point of interest. Based on the user’s point of interest, your task is to find frames in the video that contain information related to the user‘s interest and output the timestamps of these frames. Please refer to both frame description and voiceover!

\textbf{Instructions:}
\begin{enumerate}
    \item The given frames are just a part of the video, not the entire video. The entire video is about \textit{\{video\_title\}}.
    \item Don't just match relevant frames based on the keywords within the user's interest; instead, only find those that truly resonate with the user's interests in terms of content, and avoid identifying irrelevant frames.
    \item Output the timestamps of all relevant frames in the following format without any additional text or explanation (Note that the output timestamps are for video frames, not for the voiceover!):
    \[ \texttt{[frame\_timestamp1, frame\_timestamp2, ...]} \]
\end{enumerate}

------- Now, begin to analyze and output the results! ------
\begin{itemize}
    \item \textbf{User’s Interest:} \textit{\{user\_interest\}}
    \item \textbf{Frames and Voiceover:} \textit{\{frame\_voice\_list\}}
    \item \textbf{Output Timestamps:}
\end{itemize}

\subsection{Cohesive Narrative Generation}
You have access to several video segments selected based on user interest. Each segment includes a transcript from the specified time range and corresponding frame descriptions. Your task is to craft a coherent narrative using these segments.

\textbf{Instructions:}
\begin{enumerate}
    \item \textbf{Create a Narrative:} Compose a concise, coherent, and informative narrative (up to 300 words) that captures the essence of these segments. This narrative should reference both the transcript and frame descriptions. Ensure that the narrative flows naturally and avoid using exact sentences from the segments. Please also refer to the user's interest as a context.
    \item \textbf{Semantic Chunking:} After composing the narrative, divide it into semantically meaningful chunks. Each chunk should consist of one or more sentences. Do not modify or rearrange any sentences in the narrative during this process; the order of sentences must remain as you originally composed them.
    \item \textbf{Output the Narrative:} Present the full narrative and the semantically chunked version in JSON format as shown below, without any additional explanation or text.
\end{enumerate}

------- Now, let's create the narrative! ------
\begin{itemize}
    \item \textbf{User's interest:} \textit{\{user\_interest\}}
    \item \textbf{Segment information:} \textit{\{segments\}}
\end{itemize}

\textbf{Example output:}
\begin{verbatim}
{
    "overall_narrative": "The narrative you created in step 1.",
    "chunks": [
        {
            "chunk_id": 1, 
            "narrative": ""
        },
        {
            "chunk_id": 2, 
            "narrative": ""
        }
    ]
}
\end{verbatim}

\subsection{Video-centric Playback}
You have access to a narrative and the video segments from which it was composed. Each segment includes a transcript from the specified time range and corresponding frame descriptions. Your task is to rearrange the segments to best align with the provided narrative.

\textbf{Instructions:}
\begin{enumerate}
    \item \textbf{Segment Inclusion:} Ensure all provided segments are included in the final arrangement. Do not repeat segments.
    \item \textbf{Rearrange Segments:} Organize the segments in the order that best supports the narrative. Rearranging is optional if the current order already fits the narrative.
    \item \textbf{Output Format:} Submit the final arrangement in JSON format, following the example provided below.
\end{enumerate}

------- Now, let's create the narrative! ------
\begin{itemize}
    \item \textbf{Narrative:} \textit{\{overall\_narrative\}}
    \item \textbf{Segment information:} \textit{\{segments\}}
\end{itemize}

\textbf{Example output format:}
\begin{verbatim}
{
    "segments": [
        {
            "start": , # start time of this segment, do not modify
            "end": ,# end time of this segment, do not modify
            "playback_order": # the order for segment playback
        }
    ]
}
\end{verbatim}

\subsection{Narrative-centric Playback}
You have access to a narrative divided into chunks, along with the video segments with their transcripts and corresponding frame descriptions. Your task is to identify and associate the relevant segments for each narrative chunk.

\textbf{Instructions:}
\begin{enumerate}
    \item \textbf{Segment Association:} For each chunk of the narrative, identify the video segments that are most relevant based on their transcript and frame descriptions.
    \item \textbf{Chunk and Segment Requirements:}
    \begin{enumerate}
        \item \textbf{Segment Association:} Each chunk must be associated with one or more video segments that best match the content of that chunk.
        \item \textbf{Exclusive Assignment:} Ensure that each segment is assigned to only one chunk. No segment should be reused across different chunks.
    \end{enumerate}
    \item \textbf{Segment Order:} If a chunk is associated with multiple segments, arrange the segments in the order that best aligns with the narrative flow of the chunk.
    \item For each chunk in the input JSON, add a new key, \texttt{segments}, which is a list containing objects with \texttt{start} and \texttt{end} keys representing the time ranges of the associated segments. Present the results in the following JSON format.
\end{enumerate}

------- Now, let's chunk the narrative! ------
\begin{itemize}
    \item \textbf{Narrative chunks:} \textit{\{overall\_narrative\}}
    \item \textbf{Video segments information:} \textit{\{segments\}}
\end{itemize}

\textbf{Example output:}
\begin{verbatim}
{
    "chunks": [
        {
            "chunk_id": , # do not modify
            "narrative": , # do not modify
            "segments": [
                {
                    "start": , # start time of the segment
                    "end": , # end time of the segment
                },
                {
                    "start": , # start time of the segment
                    "end": , # end time of the segment
                }
            ]
        },
    ]
}
\end{verbatim}

\textbf{Note:}
\begin{enumerate}
    \item The \texttt{narrative} field must not change from the narrative chunk information as provided.
    \item Each segment only belongs to one chunk.
    \item Each chunk must have at least one segment.
    \item If you really cannot find a segment for a chunk, leave the \texttt{segments} field in this chunk empty.
\end{enumerate}

\end{document}